# FORMAÇÃO CONTINUADA: UMA PROPOSTA DE SOLUÇÃO PARA PROBLEMAS NO ENSINO DE ASTRONOMIA
*CONTINUING EDUCATION: A SOLUTION PROPOSAL FOR PROBLEMS IN THE TEACHING OF ASTRONOMY*

[1]Adriano Mesquita Oliveira*,
[2]Cibele kemeicik da silva machado,
[3]Augusto Cézar Tiradentes Monteiro,
[4]Rafael Cerqueira do Nascimento,
[5]Pollyana dos Santos.

[1]Instituto Federal do Espírito Santo. E-mail: adriano.oliveira@ifes.edu.br
[2]Secretaria Municipal de Educação de Guarapari. Email: cikemeicik@gmail.com
[3]Instituto Federal do Espírito Santo. E-mail: augusto.monteiro@ifes.edu.br
[4]Instituto Federal do Espírito Santo. E-mail: rafaelc@ifes.edu.br
[5]Instituto Federal do Espírito Santo. E-mail: pollyana.santos@ifes.edu.br

**Resumo:** O presente artigo tem como objetivo relatar e refletir sobre a experiência de aplicação de um curso de formação continuada em Ensino de Astronomia para professores das redes municipais de Laranja da Terra (ES) e Guarapari (ES). Este artigo apresenta o curso ofertado pelo Ifes-Guarapari para essas redes assim como analisa a coleta de dados extraídos dos relatos de experiência elaborados pelos cursistas que foi orientado por questões diagnósticas que permitissem a sistematização das percepções dos cursistas no processo formativo desenvolvido. Após a sistematização e análise dos textos, a reflexão foi a seguinte: a) a existência da demanda por formação na área do conhecimento astronômico; b) o potencial do Ensino de Astronomia como propiciador de práticas pedagógicas inovadoras no contexto escolar assim como seu potencial interdisciplinar; e c) apontamentos para a melhoria na construção e oferta de novos cursos de formação em Astronomia para profissionais do ensino fundamental.

**Palavras-chave**: Ensino; séries iniciais; formação continuada; astronomia; interdisciplinaridade.

**Abstract:** The goal of this paper is have bringing a reflect on the experience of applying a continuing education course in Astronomy Teaching for teachers from two municipal networks, know Laranja da Terra (ES) and Guarapari (ES). Here we are presenting a brief of the course offered by Ifes-Guarapari for these networks as well as analyzes the experience reports, font of collection of data, done by the course participants, which was guided by diagnostic questions that allowed the systematization of the course participants' perceptions in the developed training process. After the systematization and analysis of the texts, the reflection was as follows: a) the existence of a demand for training in the area of astronomical knowledge; b) the potential of Astronomy Teaching as a provider of innovative pedagogical practices in the school context as well as its interdisciplinary potential; and c) notes for improvement in the construction and offer of new training courses in Astronomy for elementary school professionals.

# 1 INTRODUÇÃO

No ano de 2018, o Campus Guarapari, do Instituto Federal de Educação, Ciência e Tecnologia – Ifes, em parceria com as redes municipais de ensino dos municípios de Laranja da Terra e Guarapari (ES), ofertou um curso de formação continuada em Ensino de Astronomia para professores e servidores da educação de modo geral. Com a proposta de qualificação e desenvolvimento profissional na área de ciências, o curso foi desenvolvido em dois municípios com dimensões distintas, regionalmente e culturalmente distantes, mas com algo em comum: as taxas de rendimento escolar e as médias de desempenho nos exames aplicados pelo Inep [1] não atingem as metas estabelecido pelo governo [2].

Nesse sentido, a proposta foi a de recorrer ao Ensino de Astronomia e suas potencialidades para a formação básica, em especial seu potencial de letramento científico e seu caráter multidisciplinar, a fim de construir e aplicar um curso de formação continuada para profissionais da educação desses municípios. O presente artigo apresenta e analisa a experiência do desenvolvimento do curso assim como avalia os dados obtidos da coleta de informações dos questionários aplicados aos professores cursistas, a fim de refletir sobre a importância da oferta de formação na área de Ensino de Astronomia e apontar caminhos para propostas efetivas de melhoria do ensino de ciências e, em particular, da Astronomia no contexto escolar.

A Base Nacional Curricular Comum (BNCC) de 2018 [3] propõe o "desenvolvimento do letramento científico" para os alunos, considerando que esse se efetue por meio da aproximação com os "principais processos, práticas e procedimentos da investigação científica". Segundo a BNCC, o processo investigativo deve estar como elemento central na formação dos estudantes e, por isso, deve ser contemplado nas situações didáticas ao longo de todo processo formativo desses. O alcance dessa proposta pode ser limitado considerando a realidade do Ensino de Astronomia. Nessa área, grande parte dos profissionais utilizam somente o livro didático como fonte de conhecimento e instrumento de ensino. A preocupação com o Ensino de Astronomia relatada em 2007 por Pinto e Fonseca [4], apresenta:

> ... em um curso em astronomia básica para professores do primeiro segmento do ensino fundamental, para que ocorresse mudança nas concepções de ensino e aprendizagem dos professores, fez-se necessário levá-los a uma situação conflitiva em que suas concepções pudessem ser questionadas, por meio de um diálogo participativo, visando a construção de um novo conhecimento.

A recorrência limitada ao livro didático mantém os profissionais alheios ao pensar científico, aos fundamentos que sustentam as discussões apresentadas, a evolução do conhecimento científico e ao atual estado da arte sobre o tema. Nesse sentido, visando possibilitar o que sugeriram Pinto e Fonseca acima, as formações continuadas possuem um papel fundamental, pois possibilitam o contato dos professores das redes de ensino com especialistas e pesquisadores da área de Astronomia e suas potencialidades.

Cabe ressaltar, que a formação continuada foi entendida como um processo de qualificação profissional que não pode ser dissociado da prática docente exercida pelos profissionais em seu cotidiano. Assim, dentre os eixos fundamentais para o desenvolvimento da formação continuada, apontados por Candau em 1997 [5], o curso de Ensino de Astronomia direciona para a valorização dos saberes docentes constituídos em sua experiência profissional assim como atenta e respeita a etapa profissional em que se encontram os professores. Com essa orientação, apesar de focar nos conhecimentos de Astronomia e suas possibilidades interdisciplinares, a formação continuada segue a interpretação de Candau, no que tange a preocupação em ser um curso voltado para a reflexão das práticas docentes. Nesse sentido, projetamos o curso de Ensino de Astronomia em função de possibilitar mudanças na experiência pedagógica, a partir de um conjunto de saberes que envolve conhecimentos em Astronomia, suas potencialidades interdisciplinares e desenvolvimento de projetos didáticos. Permitindo, assim, estabelecer uma reflexão crítica no fazer pedagógico e, a partir dele, intervir para a inserção de novas práticas e saberes em sala de aula, como proposto por Harres em 2001 [6].

> Os conceitos didáticos de professores têm sido objeto de estudos, e os resultados demonstram a relevância da influência sobre a sua prática, e constitui o conhecimento profissional. E esse conhecimento está em permanente evolução, sendo relevante que os processos de formação inicial e permanente sejam orientados para uma mudança gradativa do conhecimento. Nas formações, faz-se necessário uma postura construtivista sobre a evolução do conhecimento profissional dos professores e de seus conhecimentos prévios, para que assim também, possam adotar posturas de consideração ao conhecimento prévio de seus alunos.

Além disso, o curso de formação continuada "Ensino de Astronomia para Alunos do Ensino Fundamental" foi utilizado como instrumento de aproximação do Ifes com as redes municipais de ensino citadas anteriormente. Com isso, coube ao Ifes estruturar uma proposta que considerou, sobretudo, a nova BNCC para a área de Ciências da Natureza. Cabe destacar que nesse novo currículo nacional, uma das três unidades temáticas do ensino fundamental, "Terra e Universo", visa:

> a compreensão de características da Terra, do Sol, da Lua e de outros corpos celestes – suas dimensões, composição, localizações, movimentos e forças que atuam entre eles. Ampliam-se experiências de observação do céu, do planeta Terra, particularmente das zonas habitadas pelo ser humano e demais seres vivos, bem como de observação dos principais fenômenos celestes. Além disso, ao salientar que a construção dos conhecimentos sobre a Terra e o céu se deu de diferentes formas em distintas culturas ao longo da história da humanidade, explora-se a riqueza envolvida nesses conhecimentos, o que permite, entre outras coisas, maior valorização de outras formas de conceber o mundo, como os conhecimentos próprios dos povos indígenas originários.

Sendo assim, o curso foi direcionado a partir desse eixo temático (Terra e Universo). Nele, o Ensino de Astronomia perpassa todos os anos de formação do Ensino Fundamental, possibilitando que ele contribua não só com o desenvolvimento das habilidades investigativas do letramento científico no processo formativo dos alunos, como também, dialogue com outras áreas como a Matemática, a História, Literatura e as Artes, possibilitando a capacidade de leitura do mundo e o desenvolvimento de habilidades inerentes às outras áreas de formação.

A seguir, apresenta-se o relato da formação desenvolvida, com suas temáticas e metodologia, assim como o resultado da coleta de dados junto aos cursistas visando uma reflexão sobre a relevância do curso de formação assim os caminhos para a aplicabilidade de formações dessa natureza.

## 2 O CURSO "ENSINO DE ASTRONOMIA PARA ALUNOS DO ENSINO FUNDAMENTAL":

A primeira versão do curso de formação continuada em Ensino de Astronomia teve como público-alvo profissionais que atuam na educação básica: professores e técnicos em educação das redes de ensino de Laranja da Terra e Guarapari. Além do corpo de profissionais do Ifes responsáveis por sua execução, o curso contou com a colaboração da Ufes, via PPG-Cosmo, por meio de empréstimo de equipamentos de observação astronômica, orientações técnicas e seminários aos cursistas.

O curso ofertado contou com 105 (cento e cinco) servidores municipais matriculados. Do total de matrículas, 66% foram da PML (Prefeitura Municipal de Laranja da Terra) e 33% da PMG (Prefeitura Municipal de Guarapari). O curso teve um total de 180 h, com carga horária 50% presencial com encontros quinzenais, divididos em exposição teórica e observação do céu noturno utilizando telescópios, ambos conduzidos pelos servidores do Ifes, além de apresentação dos trabalhos desenvolvidos, conduzido pelos professores cursista e seminários de especialistas convidados da Ufes.

A partir da pergunta orientadora do curso: Você já olhou para o céu hoje? O que você vê quando olha para céu? Foram estruturados os módulos e a metodologia do curso. A Tabela 1, mostra como o curo foi dividido, a partir de uma proposta fundamentada na pedagogia de projetos e nas metodologias ativas [7-13].

**Quadro 01**: Neste quadro são apresentados os módulos, temáticas e objetivos esperados.

| Módulo | Temáticas | Objetivos |
|---|---|---|
| Tópicos de Astronomia | 1. O que você vê quando olha para o Céu? 2. Por que o Céu é azul? 3. Planetas ou Estrelas? 4. Agrupando as Estrelas. 5. Por onde caminha o Sol e os Errante? 6. A Lua 7. A influência dos astros no modo de vida da Terra. 8. Teorias Gravitacionais e modelos de Universo. 9. Como coletar informações do Cosmo? | Ampliar o conhecimento dos cursistas sobre o tema proposto e dar-lhes ferramental teórico atualizado para melhorar a qualidade de suas aulas e minimizar os erros conceituais. |
| Observações do Céu | O que você vê quando olha para o céu? | Norteados pela questão temática, incentivou-se a observação do céu noturno, como ato contemplativo, fazendo assim com que estes reestabelecessem uma conexão com a natureza e com o meio que os cerca, muitas vezes deixada de lado pela civilização moderna, permitindo que estes pudessem tirar suas próprias conclusões acerca do movimento dos astros e reconhecer alguns dos agrupamentos |

| | | | | | |
|---|---|---|---|---|---|
| | | de estrelas. | | | uma valorização dos mais velhos e um aumento da interação com os mais novos, permitindo assim a disseminação do conhecimento passado entre gerações e a fundamentação destes fundamentando-os cientificamente ou, minimamente, o apontamento das falhas. |
| Práticas astronômicas | Confecção de uma luneta | A busca pelo avanço e inovação tecnológica passa por conhecer a história e o processo evolutivo de nossa civilização dentro da temática de interesse. Neste sentido a construção de um equipamento simples e relativamente barato fez com que os cursistas tivessem em mente a complexidades da pesquisa e a dependência desta com relação a tecnologia, para melhora dos dados e aprimoramento das técnicas. | Tópicos em Matemática | Fundamentos de Geometria; | Transcrever geometricamente as estruturas e comparar com as transcrições históricas, em especial as feitas por Galileu; Compreender e fundamentar os movimentos dos astros; |
| História Oral | Como os mais velhos se relacionavam com o céu? | A busca pelo conhecimento originário e cultural, acerca da relação que os mais velhos têm com o céu, permite o resgate cultural e a inclusão da sociedade na escola, tornando-a viva e parte importante no reconhecimento e perpetuação desta cultura. Assim, tem-se | Metodologias Ativas | Aplicação de metodologia de projetos ao ensino de Astronomia | Buscar uma melhora na relação ensino-aprendizagem perpassa por testar metodologias diferenciadas. Como o objetivo aqui é tratar como uma temática, a astronomia, pode ser abordada dentro de diferentes contextos, o |

| | | uso destas metodologias não tradicionais acontece naturalmente. |
|---|---|---|

Fonte: Próprio autor.

Como trabalho de conclusão, foi solicitado aos cursistas, que se organizassem em grupos. Foram formados 17 (dezessete) no total, contando as duas redes municipais. A proposta para finalização do curso foi: (1) um relato de Experiência, detalhando as expectativas, percepções e aplicabilidade do conhecimento construído no curso em sua realidade profissional; (2) um projeto de ensino utilizando a Astronomia como ferramenta interdisciplinar. Ambos foram entregues e apresentados a uma banca examinadora.

## 3 QUANTIFICAÇÃO A PARTIR DA ANÁLISE DO RELATO DE EXPERIÊNCIA

A frequência com a qual as informações aparecem nos relatos de experiência foi usada como indicativo de relevância. Assim, elaborou-se perguntas cujas respostas eram essas informações. Assim, conseguimos fazer a quantificação a partir da análise dos relatos de experiência vivida pelos alunos, cursistas, durante o curso de formação continuada. Desse modo, conseguimos organizar as informações em cinco questionamentos, a saber: (1) O que levou o cursista a participar da formação continuada cujo tema é Astronomia? (2) Quais as potencialidades, no que tange sua aplicação em sala de aula, das metodologias utilizadas nos encontros? (3) Quais foram os pontos positivos, do ponto de vista do cursista, do curso? (4) Quais os pontos fracos do curso? (5) Como o curso contribuiu para a sua vivência profissional? (6) Quais as dificuldades encontradas no curso? Por outro lado, a contagem da frequência com a qual uma resposta surge não é trivial. A busca de elementos relevantes nos relatos de experiência esbarra, muitas vezes, no caráter interpretativo. Neste sentido, a credibilidade dos dados coletados é estabelecida pelo crivo da análise, onde evitou-se interferir na forma como o texto aparece nos relatos analisados ao passo que, também, buscamos um agrupamento das respostas próximas. Por exemplo, no trecho do relato *"...uma boa prática em sala de aula dentro do conteúdo de ciências"* (ii) que destacamos, entrou como contagem para a questão (1) em seu respectivo tópico, ou seja (1-ii), apresentados na Tabela 1. Por outro lado, no trecho, *"... foram interessantes e nos fizeram resgatar e perceber os valores inestimáveis..."* torna-se evidente que o grupo que escreveu essa frase em seu relato de experiência valoriza a o resgate da cultura local. Assim o termo em destaque foi contabilizado na questão (2) reposta (ii) como pode ser visto na Tabela 2. No trecho *"com as aulas práticas e abordagens em sala de aula, e nos despertou o interesse na elaboração de projeto a ser desenvolvido em nosso ambiente escolar"*, o destaque aqui contabilizou para (3-iv). Já no texto *"o céu que ficou coberto deu a impressão de ter sido um erro iniciar o curso."* Foi utilizado como resposta em (4-vi). Para (5-ii) foi utilizado a fala *"...uma importante contribuição para que pudéssemos enriquecer nossos conhecimentos prévios sobre ..."* e em (6-vi) utilizamos *"...todos os participantes do grupo tinham conhecimentos mínimos a respeito da temática, mas...")."* E assim, sucessivamente. Ou seja, esse padrão de análise permitiu obter os resultados apresentados nas tabelas (2-7) e sintetizadas nos gráficos (1-6). As tabelas apresentam um índice que pode variar de (i) até (ix). Com isso em mente, foi possível organizar os dados como apresentados nas tabelas de (2-7) e gráficos de (1-6).

Tabela 1: Respostas referentes à pergunta (1): O que o levou a participar da formação continuada a Astronomia?

| | Respostas | PMG | PML | Geral |
|---|---|---|---|---|
| i | Certificação | 50% | 20% | 31% |
| ii | Aprimoramento das práticas de ensino | 33% | 30% | 31% |
| iii | Conhecimentos | 17% | 20% | 19% |
| iv | Curiosidade | 17% | 0% | 6% |
| v | Tema central da BNCC | 17% | 10% | 13% |
| vi | Falta de conhecimento prévio | 17% | 0% | 6% |

Fonte: Próprio autor.

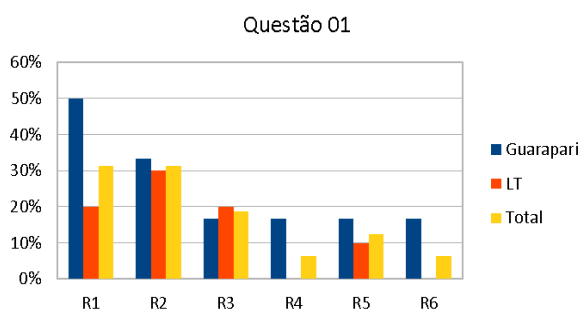

Gráfico 1: Representação gráfica dos dados expostos na tabela 1.

Tabela 2: Respostas referentes à pergunta (2): Quais as potencialidades, no que tange sua aplicação em sala de aula, das metodologias utilizadas nos encontros?

| | Respostas | PMG | PML | Geral |
|---|---|---|---|---|
| i | Desconstrução e construção do conhecimento | 33% | 10% | 19% |
| ii | Resgate da cultura e história oral | 0% | 20% | 13% |
| iii | Utilização de simuladores e modelos | 17% | 10% | 13% |
| iv | Transcrições geométricas | 0% | 20% | 13% |
| v | Pesquisas e entrevistas | 17% | 40% | 31% |
| vi | Oportunidade de mostrar o que pensava mesmo estando errado | 0% | 10% | 6% |
| vii | Trabalho em grupo | 0% | 30% | 19% |
| viii | Espaços para discussão | 0% | 10% | 6% |
| ix | Elaboração de Projetos | 17% | 20% | 19% |

Fonte: Próprio autor.

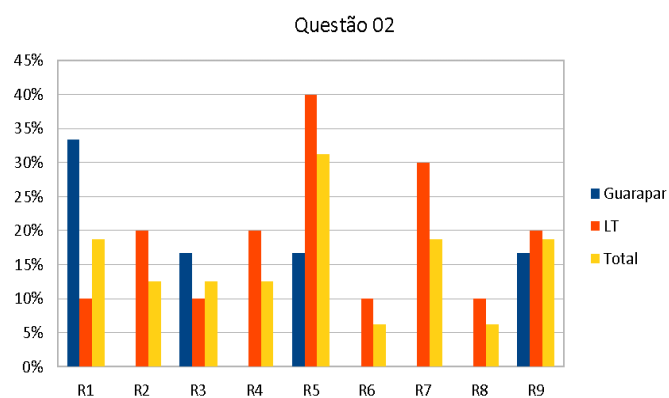

Gráfico 2: Representação gráfica dos dados expostos na tabela 2

Tabela 3: Respostas referentes à pergunta (3): Quais foram os pontos positivos da formação

| | Respostas | PMG | PML | Geral |
|---|---|---|---|---|
| i | Contribuição para atuação profissional | 33% | 10% | 19% |
| ii | Construção do pensamento científico | 17% | 10% | 13% |
| iii | Muito aprendizado (Colocaria no item 1 desse quadro) | 17% | 30% | 25% |
| iv | Novas metodologias | 33% | 30% | 31% |
| v | Despertar curiosidade | 50% | 60% | 56% |
| vi | Bons profissionais | 17% | 30% | 25% |
| vii | Resgate da cultura local | 0% | 10% | 6% |
| viii | Aulas dinâmicas | 0% | 40% | 25% |

Fonte: Próprio autor.

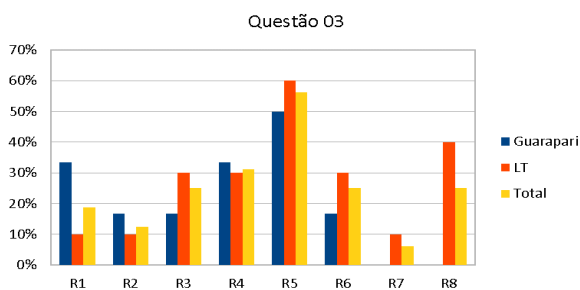

Gráfico 3: Representação gráfica dos dados expostos na tabela 3.

Tabela 4: Respostas referentes à pergunta (4): Quais os pontos fracos do curso?

| | Respostas | PMG | PML | Geral |
|---|---|---|---|---|
| i | Cansativo (Mudaria esse nome ou retiraria) | 0% | 20% | 13% |
| ii | Ausência de atividades para noites ruins | 0% | 10% | 6% |
| iii | Ociosidade durante aulas de observação | 0% | 10% | 6% |
| iv | Falta de entusiasmo durante a observação | 0% | 10% | 6% |
| v | Poucos instrumentos | 0% | 10% | 6% |
| vi | Época de aplicação do curso – Inverno | 33% | 10% | 19% |

Fonte: Próprio autor.

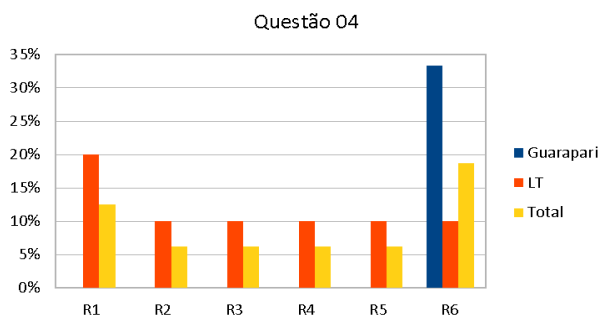

Gráfico 4: Representação gráfica dos dados expostos na tabela4.

Tabela 5: Respostas referentes à pergunta (5): Como o curso contribuiu para a sua vivência profissional

| | Respostas | PMG | PML | Geral |
|---|---|---|---|---|
| i | Trabalho diferenciado em sala de aula | 83% | 10% | 8% |
| ii | Melhorou o conteúdo para trabalhar com os alunos | 67% | 30% | 44% |
| iii | Aliar tecnologia e pensamento científico | 33% | 0% | 13% |
| iv | Tornar o aluno um cientista (Pesquisa em sala) | 0% | 20% | 13% |

Fonte: Próprio autor.

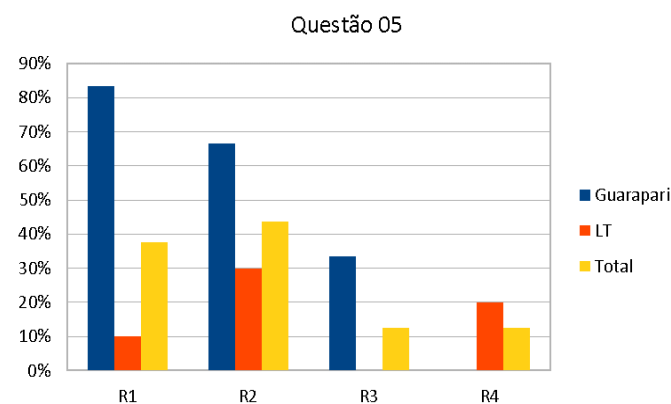

Gráfico 5: Representação gráfica dos dados expostos na tabela 5.

Tabela 6: Respostas referentes à pergunta (6): Quais as dificuldades encontradas no curso?

| | Respostas | PMG | PML | Geral |
|---|---|---|---|---|
| i | Identificação das estruturas celestes | 17% | 10% | 13% |
| ii | Construção de uma Luneta | 17% | 40% | 31% |
| iii | Atividades pesadas (Mudaria pesadas para extensas) | 0% | 10% | 6% |
| iv | Aulas (o conteúdo) de Matemática e/ou Física | 17% | 0% | 6% |
| v | Falta de conhecimento prévio sobre a | 50% | 40% | 44% |
| | temática(anterior ao curso) | | | |
| vi | Observação do céu | 50% | 20% | 31% |

Fonte: Próprio autor.

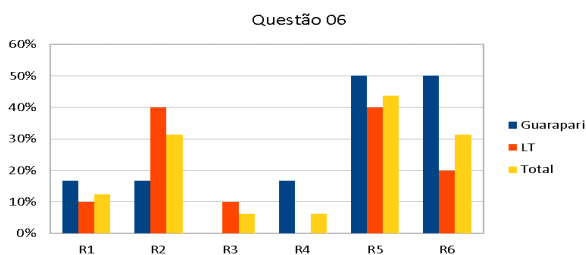

Gráfico 6: Representação gráfica dos dados expostos na tabela 6.

## 4 INDICATIVOS DA QUANTIFICAÇÃO, PERCEPÇÕES E PERSPECTIVAS

Com base no exposto, pode-se realizar algumas constatações, reflexões e apontamentos sobre a elaboração de um curso de formação continuada a partir do Ensino de Astronomia. Considerando que os dados são provenientes, especificamente, de um instrumento de coleta e na perspectiva dos cursistas, evidencia-se que as reflexões e apontamentos apresentados na seção anterior, percebe-se que para alguns cursistas a busca pela formação continuada ocorreu em função da necessidade de certificação, que é levada em consideração em editais de contratação de professores. Assim, cerca de 50% dos professores da PML citaram que esse foi um dos motivos que os fizeram ingressar no curso. Essa mesma resposta cai para 20% dos professores da PMG e corresponde a 30%, aproximadamente, do público total. Por outro lado, os dados mostram que os cursistas também se preocupam com a necessidade de buscar conhecimento e melhoria das práticas de sala de aula, essa resposta aparece em 20% e 30%, respectivamente, dos relatos. O que permite constatar que a busca pela certificação está vinculada à de capacitação, como evidencia as respostas à questão 3, nas quais identificou-se que 33% dos professores da PML citam que o curso contribuiu para a atuação profissional e uso de novas metodologias. O que, por sua vez, aponta para uma demanda de formação continuada para além do Ensino de Astronomia, mas que evidencia o quanto esse campo é capaz de contribuir com a formação pedagógica dos professores.

No tocante às práticas abordadas durante o curso, destacou-se a forma como o conhecimento trabalhado (desconstrução e construção), 1/3 dos professores de PML destacaram esse tipo de abordagem como potencialidade, enquanto 10% dos professores de PMG acharam relevante tal abordagem. Por outro lado, 1/3 dos professores desta última destacou como potencialidade o que a metodologia de projetos acarretou para o desenvolvimento do trabalho em grupo. Ainda que não tenha sido o foco da coleta, ou mesmo exista uma dificuldade em avaliar, pela participação dos professores do curso, essa discrepância pode ser reflexo de culturas institucionais distintas, existindo práticas habituais nesse sentido em Guarapari, enquanto em Laranja da Terra não.

Outro ponto que merece destaque é o da relação com o céu. O curso em Laranja da Terra possibilitou diagnosticar que no interior, a relação com o céu é aparentemente mais vivenciada e perceptível, existindo transmissão do conhecimento popular sobre esta relação. Por outro lado, na PMG os professores atuam em diversas escolas minando o contato mais próximo entre os servidores.

As questões 4 e 5 e 6 trouxeram reflexões importantes no que tange a oferta do curso e a reelaboração para novas formações futuras. Os problemas listados na questão 4 trazem apontamentos sobre o período de execução do curso e planejamento das observações do céu, fundamentais para a prática da Astronomia, considerando que 50% da formação é presencial e já conta com módulos de outras áreas, como foram os da Metodologia de Projetos, História Oral e Matemática. As aulas presenciais em alguns encontros coincidiram com as observações noturnas o que, possivelmente, gerou os dados relativos ao aspecto "cansativo" do curso. São indicadores importantes para a revisão e aplicação de futuras formações, carecendo de pesquisas junto ao público-alvo e as respectivas secretarias de educação dos municípios, a fim de que sejam integradas e equacionadas as viabilidades do Ifes como instituição ofertante e o cotidiano dos professores.

Na questão 5, constata-se que 67% da PMG e 30% da PML, o que corresponde a 44% do total de cursistas, destacou que o curso propiciou a ampliação nos conhecimentos a serem desenvolvidos no ambiente escolar. Esses dados dialogam, por sua vez, com os números que indicam a dificuldade de maior destaque (relativos à pergunta 6): 44% dos cursistas destacaram a falta de conhecimento prévio acerca dos assuntos trabalhados sobre a área de Astronomia. Somada às outras dificuldades apresentadas nas respostas, próprias das práticas da Astronomia, elas evidenciam a relevância do curso, especialmente, em relação às temáticas que envolvem o conhecimento e a prática astronômica. Estas possuem um potencial para o letramento científico, e quando limitadas aos livros didáticos durante os processos de aprendizagem, essas temáticas não surgem ou não são desenvolvidas suficientemente. Cabe ressaltar, que a perspectiva da metodologia de projetos apresentada como caminho permitiu, inclusive, apontar caminhos interdisciplinares para o desenvolvimento do Ensino de Astronomia.

Nesse sentido, os dados coletados e a vivência da aplicação da formação permitiram identificar a importância do curso. A limitação formativa, inerente a qualquer área do conhecimento, em especial na de Astronomia, leva à necessidade de propiciar a formação continuada, possibilitando aos profissionais dessas redes municipais de ensino o contato com especialistas e pesquisadores da Ufes e do Ifes. Identificou-se, assim, o que Pinto e Fonseca [4],

> pesquisas mapearam as concepções em diferentes temas relacionados à Astronomia básica e demonstraram que o conhecimento dos professores é deficitário; a falta de uma política destinada à alfabetização científica, a carência de material didático adequado, e a má formação dos professores são fatores relevantes para a causa da baixa qualidade do ensino de ciências.

Sendo assim, a construção de espaços formativos adequados que possibilitem não só o contato com conhecimentos produzidos academicamente e que acompanhem as pesquisas científicas mas também permitam o desenvolvimento de abordagens interdisciplinares que o conhecimento astronômico potencializa, são caminhos importantes para se pensar a formação continuada para o Ensino de Astronomia.

O Ifes – Guarapari criou esse espaço e diálogo com as escolas municipais de Ensino Fundamental. Assim, cabe, também, algumas considerações sobre a continuidade dos projetos que envolvem o Ensino de Astronomia, o que dá significado às formações e às reflexões sobre elas. Atualmente, o Campus Guarapari conta com um observatório astronômico e com uma nova turma de formação continuada nessa área. O observatório não existia à época de desenvolvimento do primeiro curso, objeto de reflexão desse trabalho. Sendo assim, dois aspectos novos já se apresentam como novidades oriundas dos avanços possíveis. Como demanda apresentada pelos próprios professores cursistas dessa nova turma, a visitação de algumas turmas escolares do ensino fundamental da rede pública de Guarapari ao observatório ocorreram como nova prática pedagógica desses professores. Além disso, o curso foi ampliado em seus módulos: a partir das demandas e projetos desenvolvidos no primeiro curso em Laranja da Terra, foram acrescidos os módulos de "Literatura e Universo: O cosmo como fonte de inspiração" e "Relação entre os eventos astronômicos com a vida na Terra", que no primeiro curso não teve um módulo específico, mas os assuntos tratados e as limitações dos especialistas em Astronomia acarretaram em sua inserção.

Desse modo, a reflexão apresentada acerca da aplicação de um curso de formação continuada em Ensino de Astronomia apresentou, a partir de uma pesquisa diagnóstica que ele pode ser estruturado a partir das demandas formativas apresentadas por índices educacionais ou diagnóstico de demanda por formação em determinada área, como foi o caso dos municípios de Laranja da Terra e Guarapari. No entanto, já com a segunda turma em processo de formativo, a construção de novas propostas e ofertas desse curso já apontam que podem ser orientadas a partir de diagnósticos oriundos dos contextos escolares, das preocupações e reflexões inerentes às salas de aulas e projetos escolares, dos quais surjam, também, a necessidade de redefinição das temáticas e práticas pedagógicas que não só modifiquem o modelo ofertado, mas, sobretudo, enriqueçam e contribuam com a formulação de novas práticas pedagógicas no Ensino de Astronomia durante o próprio processo de formação continuada.